\documentclass{llncs}

\usepackage[pdftex]{graphicx}
\usepackage{subfig}
\usepackage{amsmath}
\usepackage{amsfonts}
\usepackage{array}
\usepackage{multirow}
\usepackage{url}
\usepackage{subfig}

\begin{document}

\title{Does The Cloud Need Stabilizing?}

\author{
Murat Demirbas, Aleksey Charapko, Ailidani Ailijiang\\
\email {\{demirbas, acharapk, ailidani\}@buffalo.edu} \\ [2mm]
}
\institute{University at Buffalo, SUNY}

\maketitle

\begin{abstract}
The last decade has witnessed rapid proliferation of cloud computing. While even the smallest distributed programs (with 3-5 actions) produce many unanticipated error cases due to concurrency involved, it seems short of a miracle these web-services are able to operate at those vast scales. In this paper, we explore the factors that contribute most to the high-availability of cloud computing services and examine where self-stabilization could fit in that picture.

\end{abstract}

\section{Introduction}
\label{sec:intro}
The last decade has witnessed rapid proliferation of cloud computing.
Internet-scale webservices have been developed providing search services over
billions of webpages (such as Google and Bing), and providing social network
applications to billions of users (such as Facebook and Twitter).
While even the smallest distributed programs (with 3-5 actions) can produce many unanticipated error cases due to concurrency involved, it seems short of a miracle that these web-services are able to operate at those vast scales. These services have their share of occasional mishap and downtimes, but overall they hold up really well.

In this paper, we try to answer what factors contribute most to the high-availability of cloud computing services, what type of fault-tolerance and recovery mechanisms are employed by the cloud computing systems, and whether self-stabilization\footnotemark~ fits anywhere in that picture.
\footnotetext{Stabilization~\cite{dolevBook,practicalStab} is a type of fault tolerance that advocates dealing with faults in a principled unified manner instead of on a case by case basis: Instead of trying to figure out how much faults can disrupt the system's operation, stabilization assumes arbitrary state corruption, which covers all possible worst-case collusions of faults and program actions. Stabilization then advocates designing recovery actions that takes the program back to invariant states starting from any arbitrary state.}

Self-stabilization had shown a lot of promise early on for being applicable in the cloud computing domain. The CAP theorem\footnotemark~ seemed to motivate the need for designing eventually-consistent systems for the cloud and self-stabilization has been pointed out by experts as a promising direction towards developing a cloud computing research agenda~\cite{CloudAgenda2009}. On the other hand, there has not been many examples of stabilization in the clouds. For the last 6 years, the first author has been thinking about writing a ``Stabilization in the Clouds'' position paper, or even a survey when he thought there would surely be plenty of stabilizing design examples in the cloud. However, this proved to be a tricky undertaking. Discounting the design of eventually-consistent key-value stores~\cite{dynamo} and application of Conflict-free Replicated Data Types (CRDTs)~\cite{crdt} for replication within key-value stores, the examples of self-stabilization in the cloud computing domain have been overwhelmingly trivial.
\footnotetext{CAP theorem says that in the presence of a partition, it is impossible to have strong consistency and high availability~\cite{cap,capNancy,capRetroNancy}. Since cloud computing systems value high-availability (as they make money by being highly-available), weaker-consistency models such as eventual-consistency became popular.} 


We ascribe the reason self-stabilization has not been prominent in the cloud  to our observation that cloud computing systems use infrastructure support to keep things simple and reduce the need for sophisticated design of fault-tolerance mechanisms.
In particular, we identify the following cloud design principles to be the most important factors contributing to the high-availability of cloud services.
\begin{enumerate}
\item {\bf Keep the services ``stateless'' to avoid state corruption.}
  By leveraging on distributed stores for maintaining application data and on
  ZooKeeper~\cite{zookeeper} for distributed coordination, the cloud computing systems keep the computing nodes almost stateless.
  Due to abundance of storage nodes, the key-value stores~\cite{cassandra} and databases~\cite{spanner,bigtable} replicate the data multiple times and achieves high-availability and fault-tolerance.
  In our earlier work~\cite{consensus_in_the_cloud}, we explored how distributed coordination is achieved in cloud computing, and found that an overwhelming majority of cloud computing systems adopt coordination services, such as ZooKeeper, for maintaining concurrency-critical state.~\footnote{ZooKeeper is a Paxos replicated state machine implementation with a filesystem API as the interface. It is used for keeping ``metadata'' safely and consistently to the face of node failures, concurrency, and asynchrony.} 

\item {\bf Design loosely coupled distributed services where nodes are dispensable/substitutable.}  The service-oriented architecture~\cite{SOA}, and the RESTful APIs for composing microservices are very prevalent design patterns for cloud computing systems, and they help facilitate the design of loosely-coupled distributed services. This minimizes the footprint and complexity of the global invariants maintained across nodes in the cloud computing systems. Finally, the virtual computing abstractions, such as virtual machines~\cite{vm}, containers~\cite{container}, and lambda computing servers~\cite{awslambda} help make computing nodes easily restartable and substitutable for each other.

\item {\bf Leverage on low level infrastructure and sharding when building applications.}
  The low-level cloud computing infrastructure often contain more interesting/critical invariants and thus they are  designed by experienced engineers, tested rigorously, and sometimes even formally verified~\cite{awstla}. Higher-level applications leverage on the low-level infrastructure, and avoid complicated invariants as they resort to {\bf sharding} at the object-level and user-level~\cite{wikisharding}. Sharding reduces the atomicity of updates, but this level of atomicity has been adequate for most webservices, such as social networks.

\end{enumerate}

A common theme among these principles is that they keep the services simple, and trivially ``stabilizing'', in the informal sense of the term. Does this mean that self-stabilization research is unwarranted for cloud computing systems? To answer this, we point to some silver lining in the clouds for stabilization research. We notice a trend that even at the application-level, the distributed systems software starts to get more complicated/convoluted as services with more ambitious coordination needs are being build.

In particular, we explore the opportunity of applying self-stabilization to tame the complications that arise when composing multiple microservices to provide higher-level services. This is getting more common with the increased consumer demand for higher-level and more sophisticated web services. The higher-level services are in effect implementing distributed transactions over the federated microservices from multiple geodistributed vendors/parties, and that makes them prone to state unsynchronization and corruption due to  incomplete/failed requests at some microservices. At the data processing systems level, we also highlight a need for self-regulating and self-stabilizing design for realtime stream processing systems, as these systems get more ambitious and complicated as well.

Finally, we point out to a rift in the cloud computing fault model and recovery techniques, which motivates the need for more sophisticated recovery techniques. Traditionally the cloud computing model adopted the crash failure model, and managed to confine the faults within this model. In the cloud, it was feasible to use multiple nodes to redundantly store state, and easily substitute a stateless worker with another one as nodes are abundant and dispensable. However, recent surveys on the topic~\cite{taxDC,cloud_stop_computing} remark that more complex faults are starting to prevail in the clouds, and recovery techniques of restart, checkpoint-reset, and devops involved rollback and recovery are becoming inadequate.

\subsection{Outline of the rest of the paper}

To illustrate how the three design principles above are embraced in cloud computing systems, in Section~\ref{sec:arch} we present examples of services that power global-scale operations of Google and Facebook, and then delve in to the architectures of these services and their interactions with each other. We introduce the service-oriented-architecture (SOA) design pattern popular in the cloud computing software, and the microservices design pattern.

In Section~\ref{sec:faults}, we review the literature on what types of faults occur in the cloud computing systems and what type of fault-tolerance and recovery mechanisms are employed to deal with these.


In Section~\ref{sec:directions}, we point out to opportunities for applicability/adoption of self-stabilization techniques in new emerging topics in cloud computing areas, including in managing distributed coordination when composing multiple microservices to form higher level services (in Section~\ref{subsec:trans}), in dealing with complicated dataflow dependencies in realtime stream processing systems (in Section~\ref{subsec:dataflow}).

\section{Cloud Computing Design Principles}
\label{sec:arch}

\subsection{Cloud Software Infrastructure}

\begin{figure}
\centering
\subfloat[]{
\includegraphics[width=0.5\textwidth]{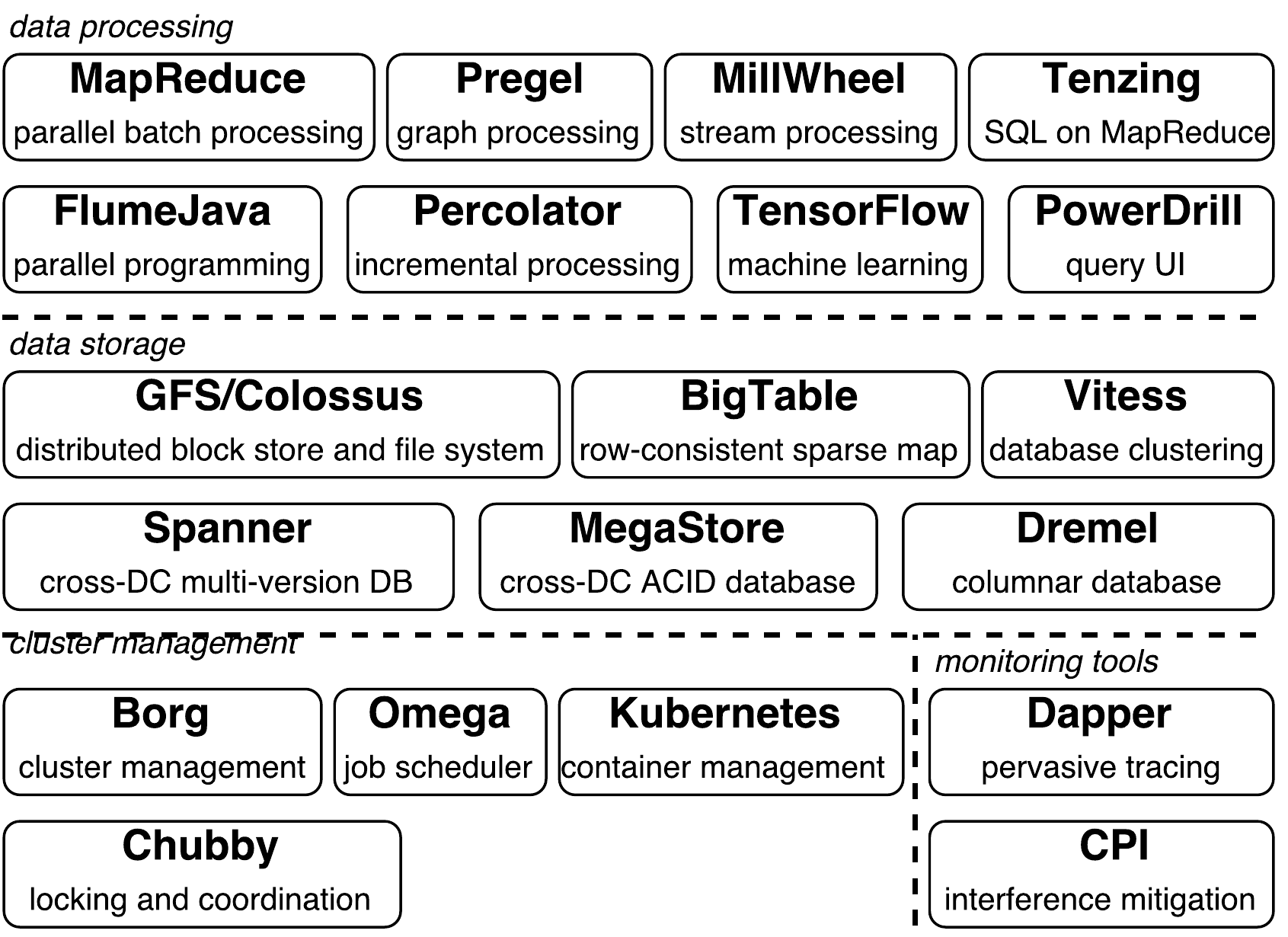}
\label{fig:google}
}
\subfloat[]{
\includegraphics[width=0.5\textwidth]{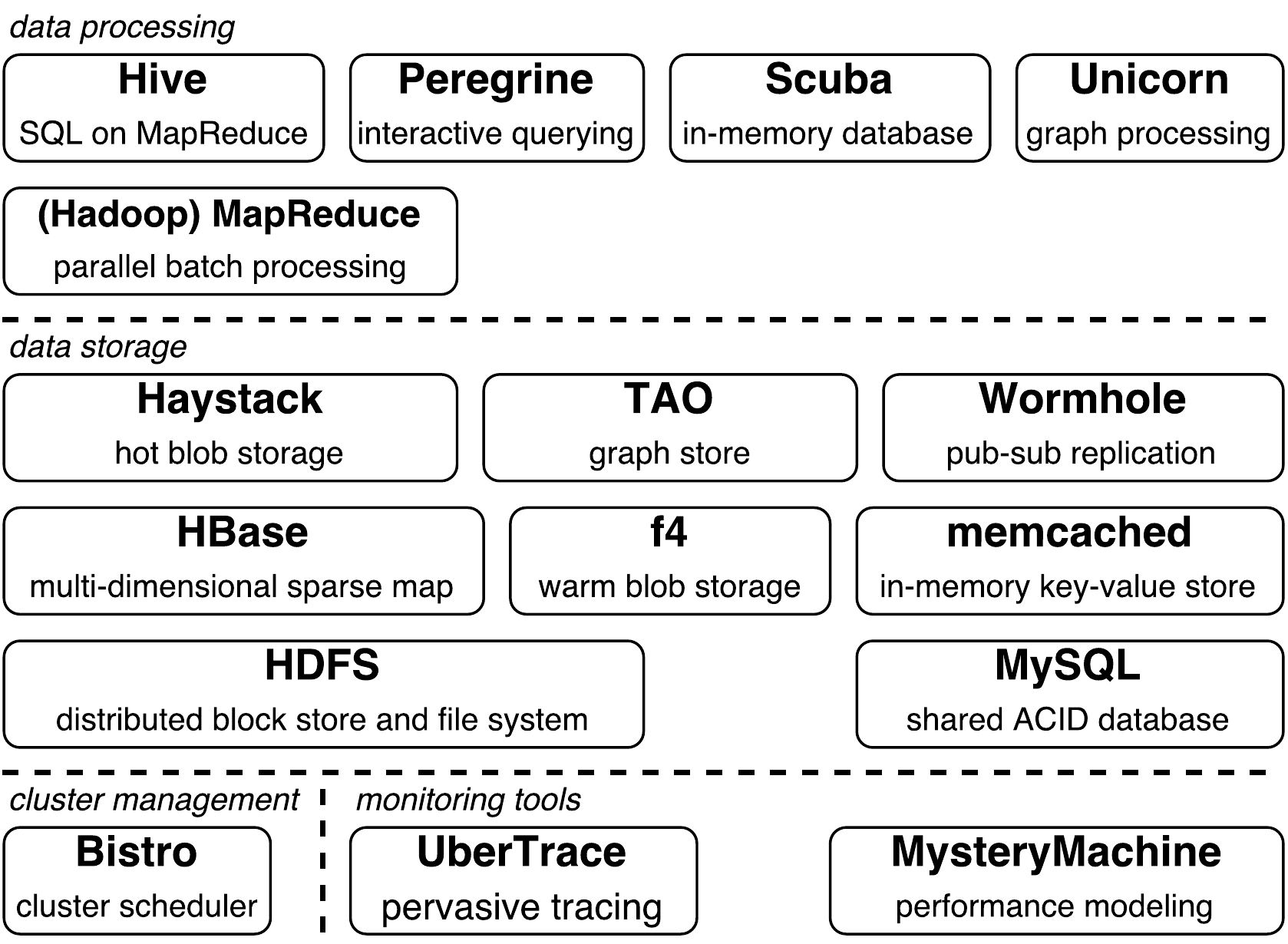}
\label{fig:facebook}
}

\caption{(a) Google and (b) Facebook software infrastructures (redrawn from \cite{schwarzkopf2015operating})}
\end{figure}

Internet scale applications such as search engine and social network demand a complicated software infrastructure with cluster management, data storage, data processing, and application layers. Each layer usually contains multiple systems to serve different use scenarios. Figures \ref{fig:google} and \ref{fig:facebook} illustrate the known systems at Google and Facebook. The cluster management layer allocates hardware resources and schedules jobs as a service provided to upper layers. Distributed monitoring systems help system admins to keep cluster healthy. There are several data storage systems that persist system state at scale and provide a rich interface with various consistency guarantees. Data processing layer implements required data analysis and representation for the applications.

Upon a closer examination, the distributed system within each layer often adopts a simple architecture pattern where the master node maintains the critical states and workers accepts tasks from the master and compute on data in a stateless manner. The master node relies on state-machine replication for fault-tolerance. Finally, these systems often employ ZooKeeper for coordination needs: including metadata management, group membership, leader election, and resource and configuration management. Cloud applications typically leverage such robust infrastructure for fault tolerance and recovery among other things.


\subsection{Service Oriented Architecture (SOA) and Microservices}

SOA is the most popular software design style in large-scale applications. In SOA, the functionality is divided into isolated services and each service can use other services through a communication protocol (e.g. RESTful) over a network. Compared to the more traditional monolithic architecture pattern, the resulting microservices architecture pattern is more scalable through functional decomposition.

Every microservice in the system is a simple, independent, distinct feature or functionality of the application. Each service instance is typically a process running on its own VM or container. Each microservice instance maintains its own state by using the data storage systems abundant in cloud computing infrastructure.
The interconnected microservices exposes a well defined REST API that is consumed by other services or clients. Representational state transfer (REST) is the most common choice for inter-service communication because it makes use of stateless protocol and standard operations (HTTP) for fast performance, reliability and the ability to grow/update without affecting the running system as a whole. Service API, therefore, is identified by its URL. Between any service requests, the client context is not stored on the server providing the service. As a result, each request from any client must contain all the information necessary to serve the request.

\subsection{Towards Stateless Computing}

\begin{figure}[t]
\centering
\includegraphics[width=0.7\textwidth]{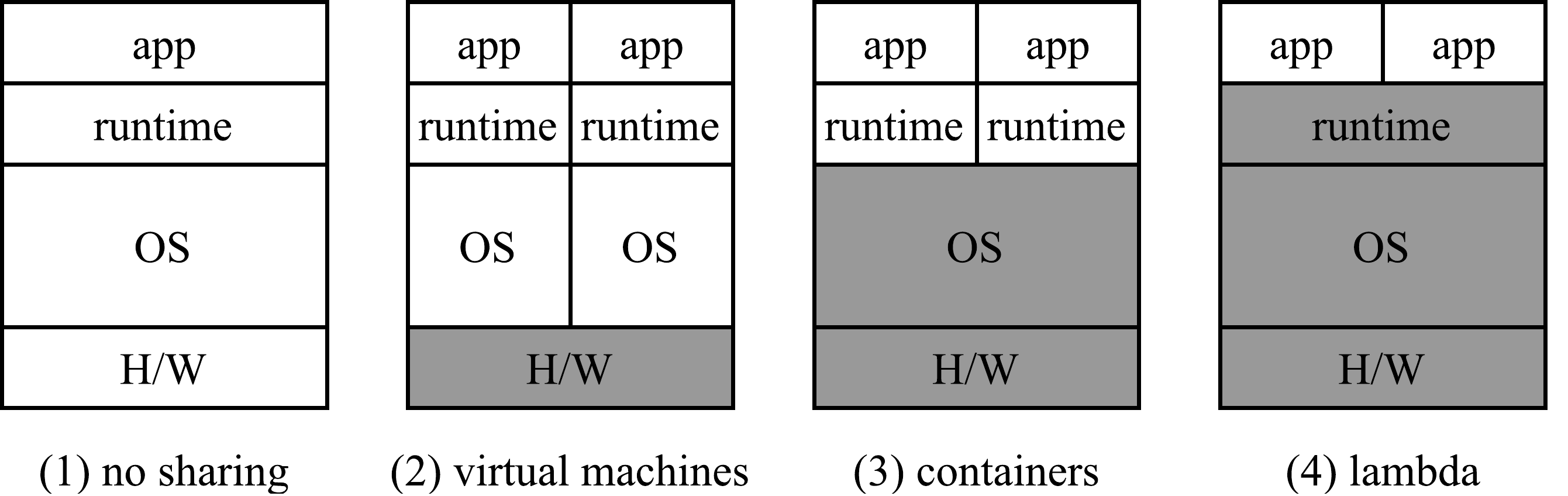}
\caption{Evolution of sharing (redrawn from \cite{serverless})}
\label{fig:sharing}
\end{figure}

The infrastructure to support the cloud application has been rapidly evolving, mostly in the opensource domain. The evolutionary trend is to increase the flexibility of cloud infrastructure while reducing the overheads via virtualization and resource sharing. Figure \ref{fig:sharing} compares these infrastructure approaches and their shared layers in gray.

The development of virtualization technologies and increased efficiency of Virtual Machines (VMs) gave rise to the proliferation of cloud computing applications. In the cloud, horizontal scalability is attained much easier by no longer requiring physical hardware to support applications. Many systems are able to run in virtual machines on the same hardware, making resource utilization higher and favoring horizontal scalability over vertical. This allowed applications to have smaller state and become more portable for the ease of quick deployment on new virtual machines. 

Containerized environments took the resource sharing beyond the hardware. Containers are isolated application environments that share hardware and an underlying operating system~\cite{container}. They are very lightweight and can be brought up online or turned off in the matter of seconds. This demanded the applications to be lightweight as well, and having large state became prohibitively expensive due to the costs of recovering the state after container migration or restart.

Finally, serverless or lambda~\cite{awslambda} computing allows multiple applications to share not only the same operating system, but also the user-space libraries and runtimes. With serverless computing, the applications are defined as a set of functions or lambda-handlers that have access to some common data-store for state. The functions themselves, however, do not carry any state from one execution to another. Handlers are typically written in a portable, interpreted languages such as JavaScript or Python. By sharing the runtime environment across functions, the code specific to a particular application will typically be small. The code size, portability and stateless nature of handlers make them inexpensive to send to any worker in a cluster.

\subsection{Coordination in the Cloud}

\begin{figure}[t]
\centering
\includegraphics[width=0.7\textwidth]{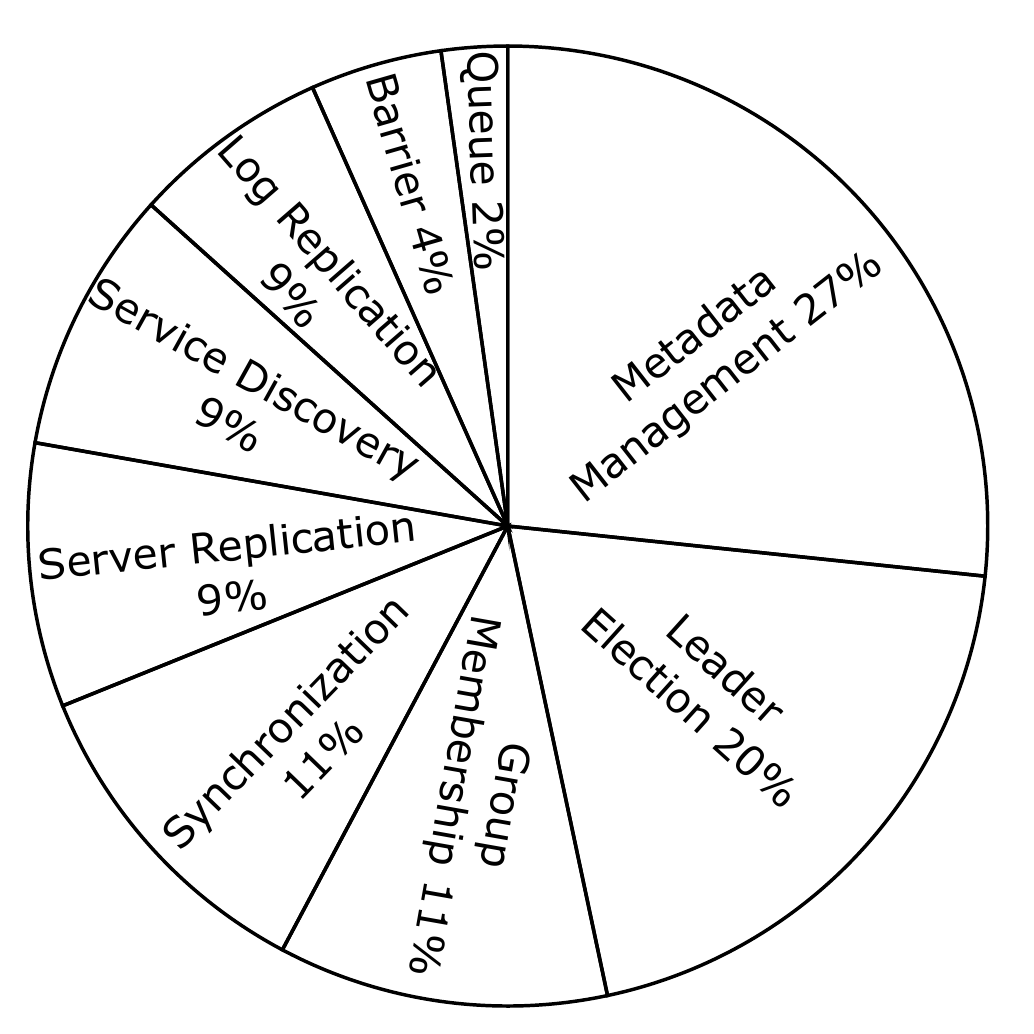}
\caption{Consensus systems usage in the cloud (redrawn from \cite{consensus_in_the_cloud})}
\label{fig:consensus_use}
\end{figure}

Coordination plays an important role in the cloud applications, especially as
the complexity of the systems grows. In our recent
work~\cite{consensus_in_the_cloud}, we surveyed the use of Paxos-like consensus
systems used in various cloud computing systems. We identified 9 distinct usecases for
the coordination as illustrated in Figure~\ref{fig:consensus_use}. The most
common of these usecases are metadata management with 27\% of the usage
scenarios surveyed, leader election (20\%), group membership (11\%), and
synchronization (11\%).


Metadata management provides a mechanism for maintaining metadata, such as configuration or state of shared objects, across the nodes in the cluster in a consistent and fault-tolerant manner. The leader election usecase allows systems to safely elect a single node acting as a dedicated master for some operations. Synchronization is another important use case that enables distributed locks. Other use cases for consensus systems in the cloud include server and log replication, barrier orchestration, service discovery and distributed queues.

\section{Cloud Computing Faults and Recovery}
\label{sec:faults}

\subsection {Cloud Computing Faults}
Failures in the cloud can be roughly categorized as small-scale faults and large-scale failures. Small faults typically only impact a few clients or some specific requests, allowing the system to process the rest of the workload. For instance, an example of such small fault would be a rejected hotel booking request due to the insufficient funds on the card. In this case, the fault does not have any impact on other users or requests, yet the system has to take a corrective action to make sure the room is not marked as reserved. The credit card failure left the booking service in a corrupted state as the room was reserved without a payment. However, this corruption is localized to a single service and not likely to spread to other services, making the room unavailability to be the only adversarial effect of the fault if left without a correction. 

Crash fault approach to fault tolerance is crucial to many of such small and localized failures. If the node failure did not impact any global state of the systems, such as persisting some corrupted data to storage, then a failure can be dealt with a simple restart or provisioning of a new node. In some cases, even if the failure is persisted to storage, it can be masked through replication and redundancy in the cloud.

More intricate small-scale failures may corrupt a state in a way that allows such corruption to propagate from one component to another. In such non-localized corruption scenarios, correcting state is more difficult, since the engineers need to account for the possibility of re-corruption cycles~\cite{leal2004scalable}. The spread of corruption across the components of the cloud system can lead to a catastrophic failure. 

Large-scale failures are faults that leave the system unavailable for significant number of users or requests. Gunawi et al. conducted a survey of catastrophic failures across various cloud applications and identified the most common reasons cloud systems crash \cite{cloud_stop_computing} as illustrated in Table \ref{tab:failure_causes}. According to their study of 597 failure reports from different systems, only 15\% of the failures are the result of some bugs in the code, while another 10\% are due to the problems in configuration. Upgrade failures, which are bugs and misconfigurations introduced during the system upgrade and maintenance account for another 16\% of failures. Many of the other failures are subject to the external events, such as security attacks, natural disasters or failure of outside dependencies

\begin{table}[!t]
\renewcommand{\arraystretch}{1.3}
\centering
\caption{Causes of Catastrophic Failures in the cloud \cite{cloud_stop_computing}}
\label{tab:failure_causes}
\resizebox{\columnwidth}{!}{

\begin{tabular}{|c|c|c|c|}
\hline
Root Cause & \# of Services & \# of Occurances & \% of Occurances\\ \hline
Unknown & 29 & 355 & - \\ \hline
Upgrade & 18 & 54 & 16 \\ \hline
Network & 21 & 52 & 15 \\ \hline
Bugs & 18 & 51 & 15 \\ \hline
Misconfiguration & 19 & 34 & 10 \\ \hline
Traffic Load & 18 & 31 & 9 \\ \hline
Cross-Service Dependencies & 14 & 28 & 8 \\ \hline
Power Outages & 11 & 21 & 6 \\ \hline
Security Attacks & 9 & 17 & 5 \\ \hline
Other Human Errors & 11 & 14 & 4 \\ \hline
Storage Failure & 4 & 13 & 4 \\ \hline
Server Failure & 6 & 11 & 3 \\ \hline
Natural Disasters & 5 & 9 & 3 \\ \hline
Other Hardware Failures & 4 & 5 & 1 \\ \hline
\end{tabular}
}

\end{table}

Cross-service dependencies are a special category of external failures originated at some service outside of the affected cloud system. These failures show that a reliability of a cloud application depends not only on the reliability of internal infrastructure and code, but also on the services external to the application. This is especially important in the context of microservices architecture, since a single unreliable microservice can cause cascading failures on its dependents.

\subsection{Recovery Models}

\textbf{Recovery by Waiting.} Recovery by waiting is typically used in cases of outside cross-service dependency failures \cite{cloud_stop_computing}. In these situations, the cloud application and infrastructure is capable of proper operation, but fails due to the unavailability of an external service, and must wait for the outside resources to become available again. As a remedy to the waiting for outside recovery, cloud applications need to design for redundancy on the external dependencies they use.

\textbf{Recovery by Restart.} Since many of the cloud components are either stateless or have small state that can be quickly replicated and loaded to new nodes, restarting the failed or slow nodes became a common practice in the cloud systems. Cloud systems are carefully guarded against server failures, making the recovery by restart an easy and safe option for handling many faults. This led to the fact that many smaller scale problems, such as memory leaks or re-optimizations done as a results of shifting workloads, are addressed by restarting the nodes instead of fixing the code or introducing additional complexity to it \cite{aws_fault_tolerant_applications}. Some cloud applications constantly induce server crashes in their infrastructure to probe the system and make sure it can handle a larger-scale faults \cite{chaos_monkey}.

However, recovery by restart is not able to address all failures happening in the cloud systems. Some failures are not caused by problems in the software, while some more intricate problems may involve distributed state corruption that can persist after individual node restarts.

\textbf{Recovery by Checkpoint Reset.} Checkpoint-reset is also a popular way to bring the system back to the correct state after the fault. These type of recovery causes the system to lose some progress and restart from some older known safe checkpoint. Because of this drawback, this type of recovery cannot be used with systems providing real-time feedback to the users. However, checkpoint-reset is a popular approach for data-processing \cite{mapreduce,naiad} and distributed machine learning \cite{tensorflow,petuum} applications.

\textbf{Recovery by Application Rollback.} Failures during the system upgrade are the single largest cause of catastrophic failures in the cloud systems \cite{cloud_stop_computing}. These failures often happen due to the bugs or misconfigurations introduced with application upgrades. The natural way to remedy such faults is rolling back to the previous stable version of the application and configuration.

Large applications, often perform gradual application upgrade or configuration rollout to the production system to minimize the impact possible bugs may bring to the systems. The idea is to catch the problem before it was deployed full-scale, thus both minimizing the severity of problem and time to roll-back. For instance, Facebook's Configerator tool controls how new configurations rollout from testing to the production environment and how the new configuration gradually expands to cover more production nodes \cite{configerator}. Configerator monitors the system behavior with every increase of new configuration's coverage and can automatically rollback to the previous good configuration if the problem is detected during the rollout.

\textbf{Recovery by State Repair.} An application state can be corrupted due to bugs in the code or configuration, however not every cloud application can tolerate being reset to a prior checkpoint or some predetermined safe state. For such application, engineers must design a corrector \cite{detectors_correctors} that fix the application state when a corruption occurs. Many severe bugs are capable of corrupting the state of the application. For instance, Leesatapornwongsa et al. compiled the database of known distributed concurrency (DC) bugs, one of the most complicated bug types in distributed systems from popular cloud-scale applications \cite{taxDC}. The authors also analyzed the fixes that were devised to address these issues in the code. 

Many DC bugs in \cite{taxDC} were fixed in an offline manner by patching the code to prevent the faulty execution from occurring, but some more intricate bugs could not be addressed in such manner. Unlike the preventable bugs caused by bad message timing or incorrect timing assumptions, the more difficult DC bugs were caused by unfortunate and uncontrollable timing of the component failures in the cloud applications.  Instead, such bugs were fixed by adding a simple corrector to the system that forward corrects the global state to a consistent state while the system is running.  

\textbf{Offline Recovery.} Many of the DC bug fixes mentioned earlier operated at the code level, preventing the problem from occurring in the fixed version. These fixes happen offline, as the application needs to go through a development cycle of making a fix, testing and deployment.
In other words, the engineers develop the correctors in an ad-hoc manner on a case-by-case basis.

\section{Future Directions for Stabilization in the Cloud}
\label{sec:directions}

\subsection{Distributed Coordination at Application Level}
\label{subsec:trans}
Cloud applications are often build on top of frameworks and architectures that hide the complexity of distributed systems. For example, MapReduce provides powerful batch-processing capabilities while keeping a simple API and allowing the application developer to write code as if they were developing a single-threaded application \cite{mapreduce}. Distributed transactions accomplish similar goals, as they allow developers to think of distributed and concurrent requests as sequential operations on a single machine. 

\begin{figure}[!t]
\centering
\includegraphics[width=2.5in]{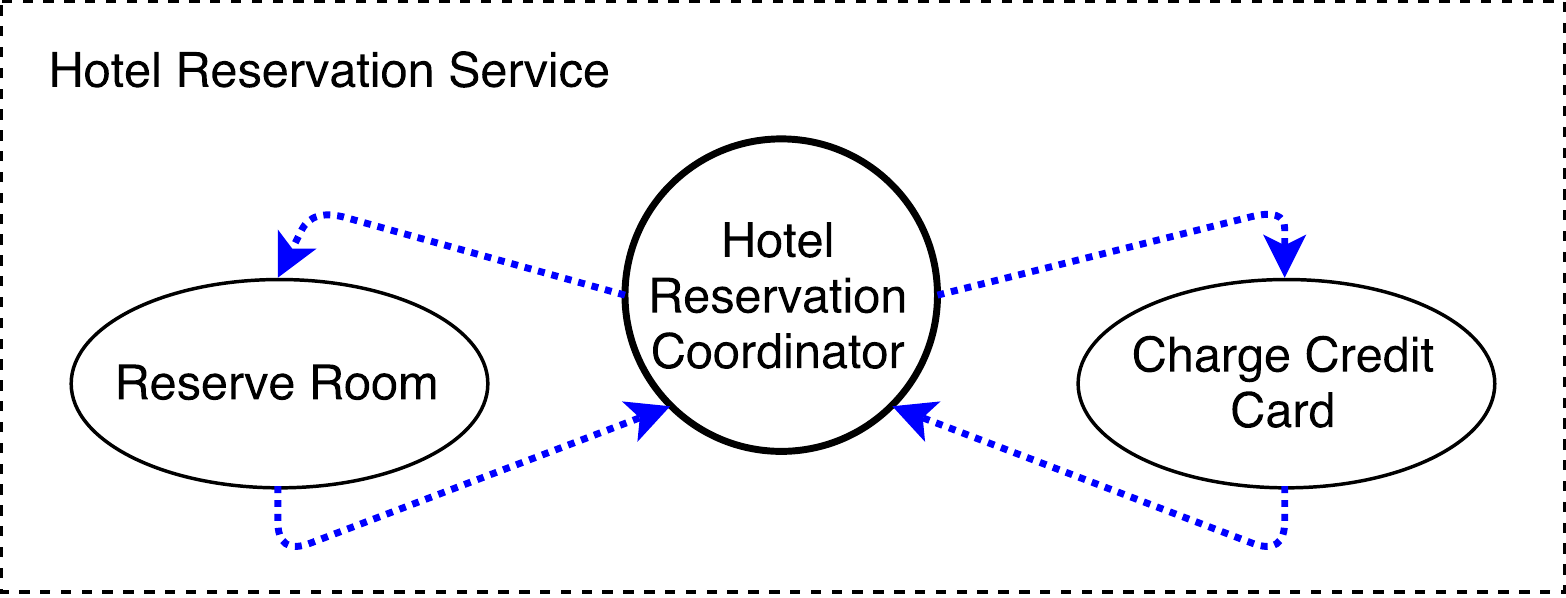}
\caption{An example of a monolithic hotel reservation service.}
\label{fig:hotel_reservation1}
\end{figure}

A transaction updates the state of multiple remote nodes or components in the cloud application in such a way that either all components are successful at changing their state or none. For example, consider our earlier hotel reservation application that perform multiple actions for the booking to be complete. The application consists of a coordinator and two components: room reservation and credit card processing, as shown in the Figure \ref{fig:hotel_reservation1}. With the help of transactions, the application can tolerate component failures without the global state being affected. For instance, if the room reservation component successfully marks the room as reserved, but the credit card processing fails, then the system will abort the transaction and no room-reservation will be preserved in the state of the application. 
Traditionally, transactional systems \cite{spanner,sinfonia} employ atomic commit protocols, such as two-phase commit (2PC), to perform transactions. However, 2PC is regarded as slow protocol with a number of corner cases that affect its performance and availability \cite{pnuts,bigtable}.

\begin{figure}[!t]
\centering
\includegraphics[width=2.5in]{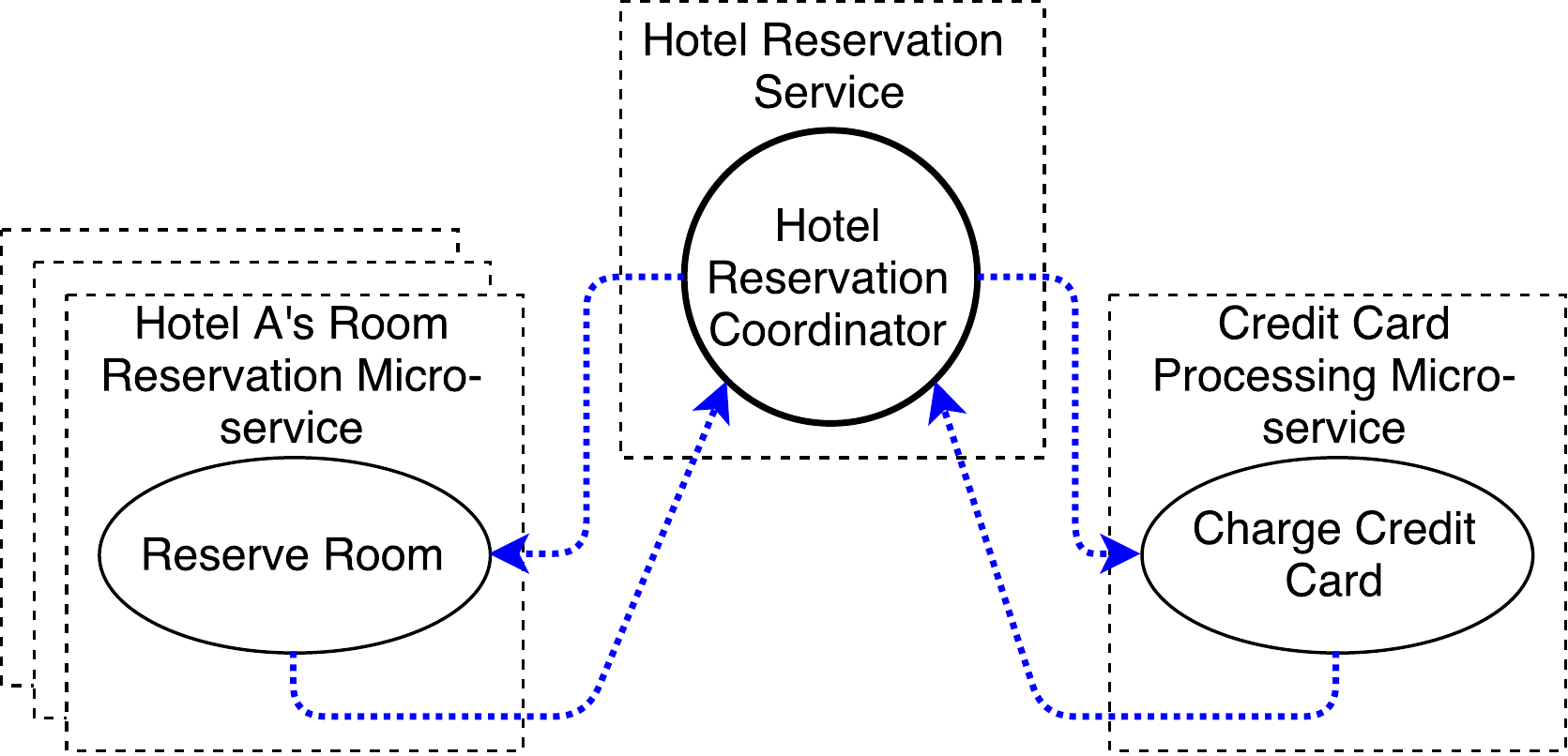}
\caption{An example of a micro-service hotel reservation system.}
\label{fig:hotel_reservation2}
\end{figure}

Transactions do not fit well in the geodistributed heterogenous cloud systems due to performance reasons~\cite{apostateHelland}. Moreover, developers cannot always rely on having a common coordination tool in microservice designs, because some of the services may be external to the application and maintained by another party. Returning to our hotel booking example, every component of the reservation service had little knowledge of other components' states, but at least these components were part of the same infrastructure and could coordinate. Now consider an example in which the hotel reservation system needs to interact with different external booking microservices for each of the hotel brands, as illustrated in Figure~\ref{fig:hotel_reservation2}. An external microservice can mutate its own state, such as booking a room or canceling the reservation, but it cannot participate in the internal transactions, since the infrastructure and protocols to coordinate between the microservices are lacking.

\begin{figure}%
\centering
\subfloat[][]{\includegraphics[width=2.22in]{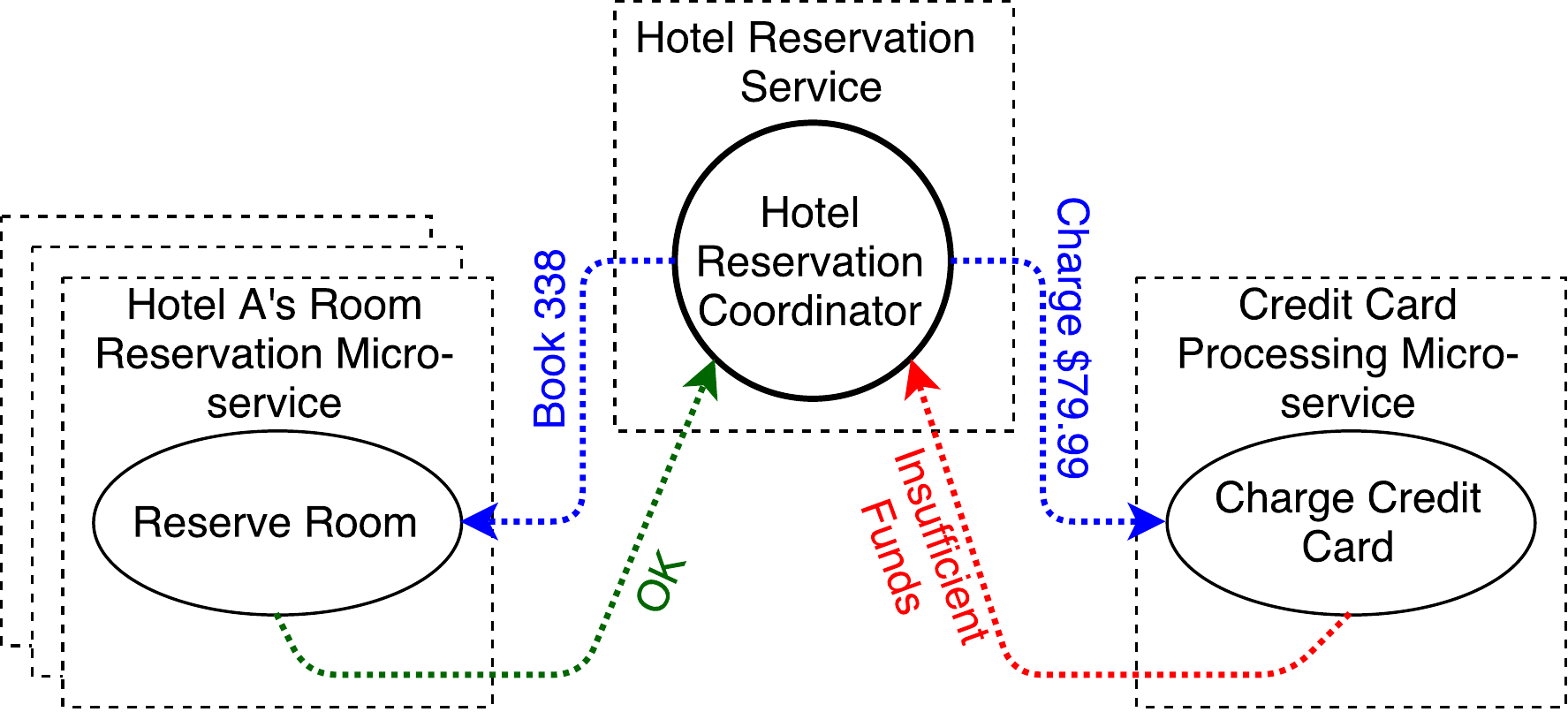}}%
\qquad
\subfloat[][]{\includegraphics[width=2.22in]{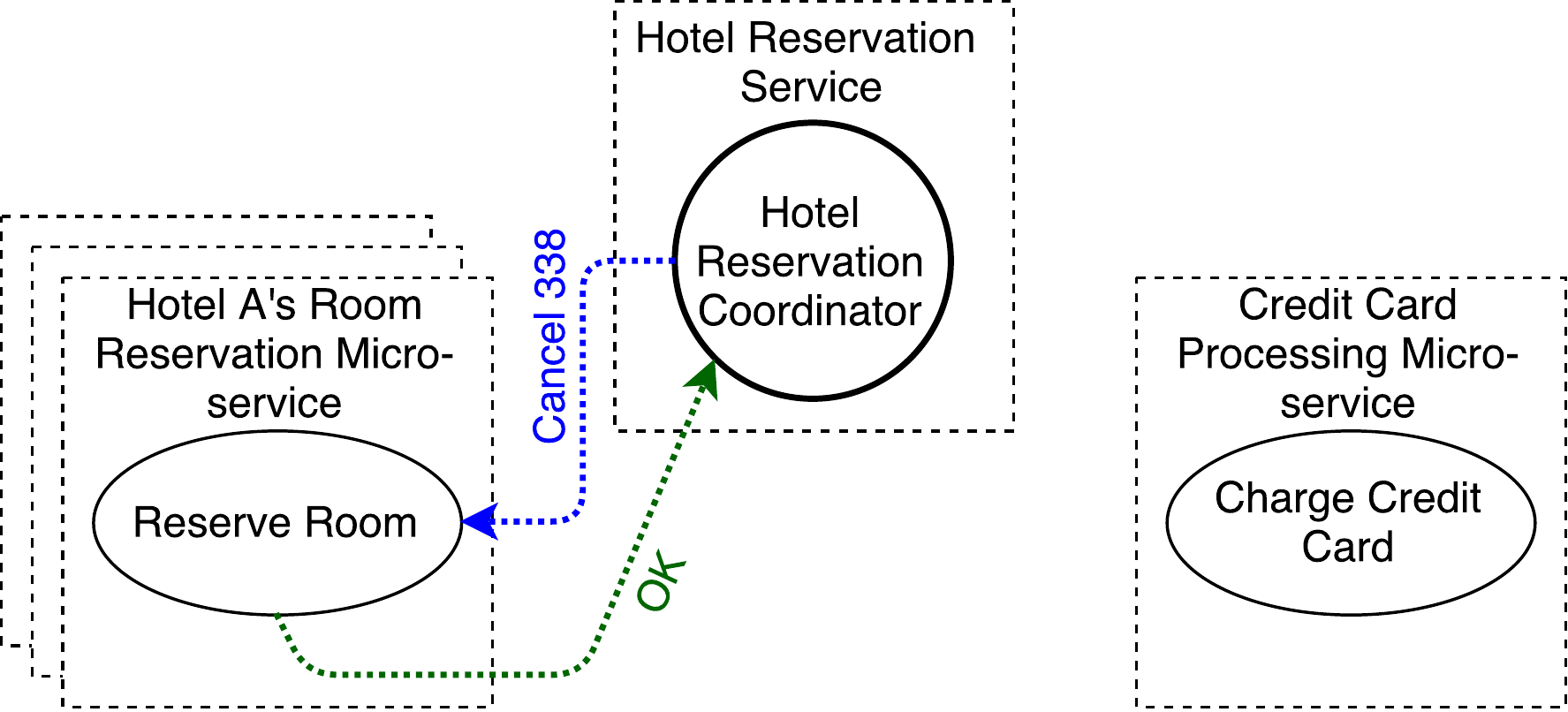}}
\caption{Hotel reservation example with corrective actions. (a) Credit card micro-service is not successful at charging the card. (b) Correcting the state of room-booking micro-service.}
\label{fig:corrective}%
\end{figure}

Despite the fact that transactional solutions do not work well with microservices architectures especially in the presence of outside actors, we often need to provide some of the transactional guarantees to such cloud systems. In particular, if one of the microservices in the request is not successful, we need to revert the state of the microservices that have already changed their states. We illustrate this process in the Figure \ref{fig:corrective}.
These corrective actions are typically written at the coordinator layer of the application in an ad-hoc manner and are not enforced by some specialized protocol. With this corrective approach, engineers need to provide undo-actions for the anticipated failures. For instance, if the room booking micro-service has already reserved the room when the credit card processing micro-service failed to charge the credit card, the coordinator needs to issue a corrective action to the booking microservice and cancel the reservation in order to bring the overall global system in a consistent state.

Self-stabilization can be useful in these scenarios,
where distributed microservices need to be coordinated and kept eventually consistent to the face of failures.
On the other hand, in order to become applicable for large scale cloud systems, self-stabilization methodology should allow certain compromises. For instance, recovering to an approximate invariant can be supported. Another example would be to design stabilization in a limited/specialized framework written to handle distributed coordination, such as distributed sagas~\cite{distributed_saga}.

\subsection{Stabilization in DataFlow Systems}
\label{subsec:dataflow}

Modern data-processing~\cite{naiad,heron} and distributed machine learning systems \cite{tensorflow} often represent the computation as a graph with data flowing along the edges and vertices representing the computations. These systems tend to utilize checkpoint-reset as their failure recovery model, however this may be inadequate for online stream processing systems, such as Twitter’s Heron~\cite{heron}. Under checkpoint reset, an online stream-processing system would lose all computations happening after the checkpoint in case of a recovery. Heron runs on a large number of small-state workers that receive their tasks from a centralized streaming manager and utilizes recovery by restart model for many faults, as the small tasks can be rescheduled to different workers as needed. 

Dhalion is a system that aims to automate many of the actions undertaken by Heron’s engineers for fault recovery and performance tuning~\cite{dhalion}. Dhalion achieves this through monitoring of many vital parameters and using the monitoring data to detect abnormalities in operation and come up with a set of possible causes for abnormal behavior. Dhalion uses the probable causes to identify the best possible fix out of the ones available to it and apply that correction to the systems, bringing it to a healthier state.  The systems attempts to resolve the problems at the higher, infrastructure level instead of the application-logic level. For instance, if Dhalion detects the performance is inadequate, it will attempt to scale up by adding more workers. Similarly, if Dhalion sees a faulty worker, it will attempt moving the tasks away from that worker into a healthy one. 

\section{Concluding remarks}
\label{sec:concl}

In this paper, we presented the design principles that contribute most to the high-availability of cloud computing services, and reviewed the type of fault-tolerance and recovery mechanisms employed by the cloud computing systems. We argued that since cloud computing systems use infrastructure support to keep things simple and reduce the need for sophisticated design of fault-tolerance mechanisms, self-stabilization has not been prominent in the cloud. Finally, we identified recent trends in cloud computing that motivate the design of more sophisticated fault-tolerance mechanisms/frameworks, namely the implementation of distributed coordination over microservices, the development of sophisticated realtime dataflow processing systems, and the increased frequency of intricate failures due to complicated interactions of services.


In order to adopt self-stabilization to address availability issues in cloud,
we believe that it is important to adhere to the simplicity and ``infrastructure-first'' principles in the cloud computing systems.
We anticipate that the resulting solutions would make compromises to be ``practically self-stabilizing''\cite{practicalStab} rather than ``traditionally/formally self-stabilizing''. For example,
while ``arbitrary state corruption'' model is interesting theoretically,
it is infeasible to realize at the implementation level, and 
it is overkill in that the return on investment is unjustified.
We belive that there is a need to define restricted corruption models and healing techniques that leverage existing cloud infrastructure. One may consider allowing only noncritical variable corruption as in~\cite{elementSecurity} by leveraging the cloud infrastructure (such as ZooKeeper and key-value stores) to protect precious variables. Or an alternate model is to allow only specification/API-level variable corruption (reasoning those are the state that matter) and deal with implementation state corruption by leveraging cloud infrastructure.

\bibliographystyle{acm}
\bibliography{ailidani,acharapk,murat}

\end{document}